\documentclass[conference]{IEEEtran}
\ifCLASSINFOpdf
\else
\fi
\usepackage{graphicx}                 
\usepackage{amsmath,amssymb,amsfonts} 
\usepackage{hyperref}                 
\usepackage{array}
\begin{document}
%
\title{Coded Adaptive Linear Precoded Discrete Multitone Over PLC Channel}

\author{\IEEEauthorblockN{Fahad Syed Muhammmad\IEEEauthorrefmark{1}, Jean-Yves Baudais, Jean-Fran\c cois H\'elard, and Matthieu Crussi\`ere}
\IEEEauthorblockA{Institute of Electronics and Telecommunications of Rennes\\
35043 Rennes Cedex, France\\
\IEEEauthorrefmark{1}Email: fahad.syed-muhammad@ens.insa-rennes.fr}
}


%



\maketitle
\begin{abstract}
Discrete multitone modulation (DMT) systems exploit the capabilities of orthogonal subcarriers to cope efficiently with
narrowband interference, high frequency attenuations and multipath fadings with the help of simple equalization filters. Adaptive linear precoded discrete multitone (LP-DMT) system is based on classical DMT, combined with a linear precoding component. In this paper, we investigate the bit and energy allocation algorithm of an adaptive LP-DMT system taking into account the channel coding scheme. A coded adaptive LP-DMT system is presented in the power line communication (PLC) context with a loading algorithm which accommodates the channel coding gains in bit and energy calculations. The performance of a concatenated channel coding scheme, consisting of an inner Wei's 4-dimensional 16-states trellis code and an outer Reed-Solomon code, in combination with the proposed algorithm is analyzed. Theoretical coding gains are derived and simulation results are presented for a fixed target bit error rate in a multicarrier scenario under power spectral density constraint. Using a multipath model of PLC channel, it is shown that the proposed coded adaptive LP-DMT system performs better than coded DMT and can achieve higher throughput for PLC applications.\\
\end{abstract}


%
\IEEEpeerreviewmaketitle

\section{Introduction}
The thriving growth of indoor and outdoor networks is driving an ever increasing demand for high speed data transmission. Among many competing technologies, power line communication (PLC) has its unique place due to already available power supply grids in both indoor and outdoor environments. Harsh channel characteristics with deep fades caused by multipaths, frequency-dependent cable losses, and hostile noise conditions are some of the biggest hurdles in the way of PLC system designers. To cope with them, one requires robust and efficient modulation and channel coding techniques.

Combinations of multicarrier (MC) and linear precoding (LP) have proved their significance in the digital subscriber line (DSL) context~\cite{mallier}. The idea of using linear precoding to improve performance over fading channels is related to that of~\cite{boutros},~\cite{hero} and~\cite{rainish}, where a real orthogonal precoder is applied to maximize the channel cutoff rate~\cite{rainish}, or maximize the minimum product distance~\cite{boutros}, \cite{hero}. Uncoded linear precoded discrete multitone (LP-DMT) has already been suggested for PLC networks in~\cite{matthieu1} and \cite{matthieu2} with a loading algorithm that handles the subcarrier, code, bit, and energy resource distribution among the active users but without taking into account the channel coding scheme. Assuming perfect channel state information (CSI) at the transmitting side, energy and bits are efficiently distributed among the precoding sequences by the loading algorithm to achieve either high throughput or high robustness. In this paper, we examine the performance of an LP-DMT system exploiting a resource allocation algorithm which takes into account the channel coding scheme. The loading algorithm, proposed in~\cite{matthieu1} and ~\cite{matthieu2}, is modified to accommodate the coding gains associated to the channel coding scheme. The proposed bit and energy allocation algorithm can be used in combination with any channel coding scheme, no matter it has constant or variable coding gains for different modulation orders, provided the obtained coding gains are known for all the modulation orders. 

Given an adaptive LP-DMT system, the suitable coding scheme should have large coding gains, reasonable implementation complexity and some measure of burst immunity. Selected on these bases, the proposed concatenated channel coding scheme consists of an inner Wei's 4-dimensional (4D) 16-states trellis code~\cite{wei} and an outer Reed-Solomon (RS) code. This combination has already proved its significance in xDSL systems and has been included in many standards~\cite{ituvdsl2}. The efficient performance of Wei's 4D 16-states trellis code over gaussian channel has also been demonstrated in~\cite{transcioffi} and~\cite{confcioffi}. Here, we analyze the performance of the proposed concatenated channel coding scheme for an LP-DMT system using a multipath reference model of power line channel~\cite{zimmermann}.

The rest of the paper is organized as follows. In Section \ref{sysmod}, an uncoded LP-DMT system is presented along with a brief description of the chosen coding scheme and the structure of a coded LP-DMT system. The modified bit loading algorithm is described taking into account the coding gains obtained from the channel coding scheme. Section \ref{theory} presents the theoretical derivations related to the coding system and gives the expression of the obtained coding gain along with the loss incurred due to redundancies. In Section \ref{simres}, simulation scenarios are discussed and results are presented for the classical discrete multitone (DMT) system and the proposed adaptive LP-DMT system using a multipath PLC channel model~\cite{zimmermann} for both coded and uncoded implementations. Energy distributions for coded DMT and coded LP-DMT are also compared. LP-DMT and DMT simulations, for both coded and uncoded scenarios, are run for various channel lengths, while using length profiles of the power line channel model suggested in~\cite{zimmermann}. It is shown that the proposed coded LP-DMT system with the modified bit loading algorithm performs better than coded DMT and achieves higher throughput for PLC applications. Finally, Section~\ref{conc} concludes this paper.

\section{System Model}\label{sysmod}
\subsection{Adaptive LP-DMT}

\begin{figure}[!t]
\includegraphics[scale=0.273]{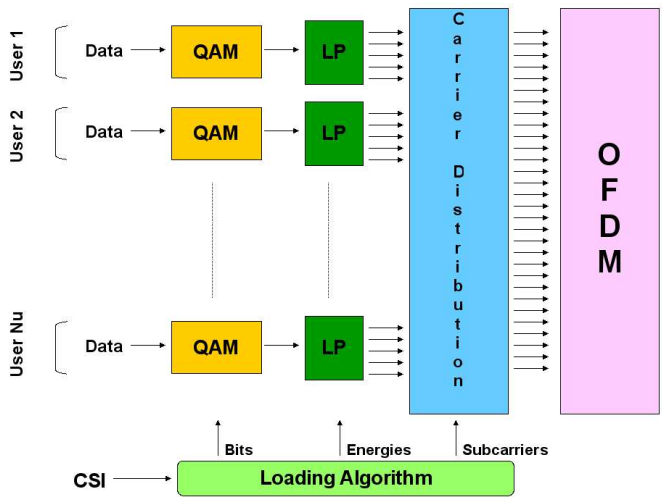} 
\caption{Uncoded LP-DMT transmitter structure}
\label{fig1}
\end{figure}

The structure of the considered adaptive LP-DMT system is shown in Fig.~\ref{fig1} as suggested in~\cite{matthieu1}. The entire bandwidth is divided into $N$ parallel subcarriers which are split up into $N_k$ sets `$S_{k}$' of $L_c$ subcarriers. The precoding function is then applied block-wise by mean of precoding sequences of length $L_c$. In fact, the precoding function can be viewed as a spreading component carried out in the frequency domain as in multicarrier code-division-multiple-access (MC-CDMA). Factor $L_c$ is such that $L_c \ll N$, which implies that $N_{k} = \lfloor\frac{N}{L_{c}}\rfloor$. Note that the subsets in a given set are not necessarily adjacent. Each user $u$ of the network is being assigned a set $B_{u}$ of subsets $S_{k}$. We emphasize that $\forall{u},~B_{u}$ are mutually exclusive subsets.
Consequently, multiple access between the $N_{u}$ users is managed following a frequency division multiple access (FDMA) approach, instead of a code division multiple access (CDMA) approach. It is worthy to mention here that we are going to consider the only case of a single user multiple block system which can be easily extended for a multi user multiple block scenario. The number of precoding sequences used to spread information symbols on one subset $S_{k}$ is denoted by $N_{c}^{(k)}$, with $0 \leq N_{c}^{(k)} \leq L_c$ since we assume orthogonal sequences. A certain amount of energy $e_{i}^{(k)}$ will be assigned to each precoded sequence $c_{i}$ associated to a given modulation symbol of $b_{i}^{(k)}$ bits, where $1 \leq i \leq N_{c}^{(k)}$. Finally, in the following, $H = \{1,\ldots,N\}$ will be the set of the useful subcarriers of the multicarrier spectrum.
\subsection{Applied Channel Coding Schemes}
The suggested coding scheme for adaptive LP-DMT system consists of an inner Wei's 4D 16-states trellis code and an outer RS code.\\
\subsubsection{Wei's 4D 16-States Trellis Code}
Trellis coded modulation combined with interleaving enables a better trade off between performance and bandwidth efficiency, while enjoying low-complexity Viterbi decoding. Trellis coded modulation systems achieve significant distance gains which are directly related to the number of states. However, the coding gain saturates upon approaching a certain number of states and the constellations must be changed to achieve higher gains. Multidimensional constellation then gives a potential
solution. An inherent cost of 2D coded schemes is that the size of the constellation is doubled over uncoded schemes. This is due to the fact that a redundant bit is added to every signaling interval. Without
that cost, the coding gain of those coded schemes would be 3 dB greater. Using a multidimensional constellation with a trellis code of rate $\frac{m}{m+1}$ can reduce that cost because fewer redundant bits
are added for each 2D signaling interval. For example, that cost is reduced to about 1.5 dB if
four-dimensional constellations are used, which is the case in the suggested coding scheme. The trellis code considered here is a 4D 16-states code developed by Wei~\cite{wei}. This code provides a fundamental coding gain of $\gamma_{f,dB} =$~ 4.5 dB, computed as a 6.0~dB increase in the minimum squared distance between allowable signal sequences, less a 1.5~dB penalty incurred for a normalized redundancy of 0.5 bits per 2D symbol~\cite{confcioffi}.\\
\subsubsection{RS Codes}
The binary data at the input is first fed to an outer interleaved RS code with code length $n$ and information length $k$. To correct $t$ random errors in a block of $n$ symbols, $n - k = 2t$ parity check symbols are required for an RS code. The RS code used here is based on a finite field (also known as Galois Field) GF($2^8$), and can have 256 different values between 0 and 255. It is a shortened RS code RS(240,224), supported in many standards~\cite{etsivdsl}, and can correct up to 8 erroneous bytes.
\subsection{Coded Adaptive LP-DMT}
\begin{figure}[!t]
\begin{center}
\leavevmode
\includegraphics[scale=0.224]{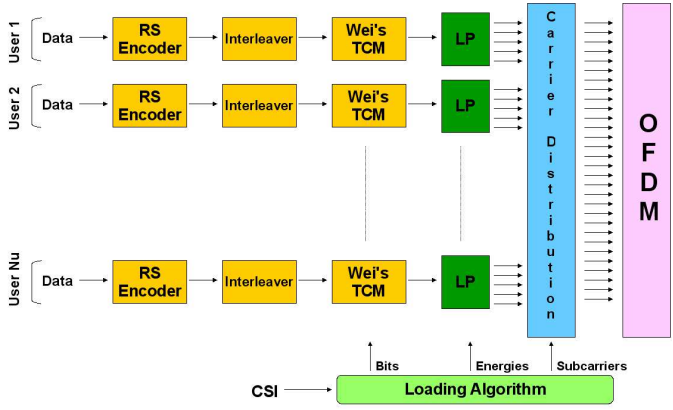} 
\caption{Coded LP-DMT transmitter structure}
\label{fig2}
\end{center}
\end{figure}

The structure of the suggested system, including the proposed channel coding scheme, is shown in Fig.~\ref{fig2}. Wei's trellis code operates on the bits allocated to the precoding sequences and produces two 2D points at its output.
In Fig.~\ref{fig2}, only a single output is shown for Wei's trellis encoder, because both 2D outputs are allocated to the same code. Also multiple copies of Wei's encoder is shown for the purpose of illustration whereas, in practice a single encoder is used to encode across the precoding sequences as discussed in~\cite{transcioffi}-\cite{confcioffi}, where a single encoder is used to encode across the subcarriers. It will be shown in Section \ref{simres} that similar to independent and memoryless subchannels in a DMT scenario, precoding sequences are also independent and memoryless in an LP-DMT scenario. The gain obtained from the application of trellis code will therefore be the same as that obtained in an intersymbol interference (ISI) free environment.
On the other hand, the RS code operates on the binary stream at the input of the system, before the bit allocation block as shown in Fig.~\ref{fig2}. A convolutional interleaver is used to spread the errors over a number of RS codewords.
\subsection{The Loading Algorithm}
The bit loading algorithm presented here is a simplified form of that presented in~\cite{matthieu1} because single user multiple block scenario is considered. The modified algorithm takes into account the channel coding scheme and is modified to accommodate the coding gains obtained from the coding scheme. Strictly speaking, all the subcarriers are allocated to
the only user in this case. The achievable data rate on a given subset $S_k$ at a given target symbol error rate (SER) can be written as

\begin{equation}\label{eqn-1}
R_{k} = L_c \log_2 \left( 1 + \frac{1}{\Gamma} \frac{L_c}{\sum\limits_{n\subset S_k} \frac{1}{|h_n|^2}} \frac{E_s}{N_0}\right)
\end{equation}
where $|h_n|^2$ is the gain of subchannel $n$, $\Gamma$ is the normalized signal to noise ratio (also known as SNR gap), $E_s$ is obtained from a given power spectral density (PSD) constraint and $N_0$ is the additive background white Gaussian noise level. SNR gap, $\Gamma$, has a constant value for all the modulation orders of uncoded quadrature amplitude modulation (QAM) for a fixed target SER. However, (\ref{eqn-1}) does not provide any practical throughput because it assumes infinite granularity.\footnote{infinite granularity: non-integer modulation order} The task here is therefore to find an appropriate bit distribution for $L_c$ available codes of each subset $S_k$, which maximizes the data rate. In the following, we discuss the allocation policy that handles the finite granularity problem.

Given an SNR gap $\Gamma$, a precoding factor $L_c$ and a transmission level $E_s$, the rate achieved by an adaptive LP-DMT system using discrete modulation is maximized if, on each subset $S_k$, $b_i^{(k)}$ bits are allocated to codes $c_i$ and $b_i^{(k)}$ is given as
\begin{equation}\label{bik}
b_i^{(k)} = 
\begin{cases} \lfloor R_k/L_c \rfloor +1 & (1 \leq i \leq n_c^{(k)})
\\
\lfloor R_k/L_c \rfloor & (n_c^{(k)} < i \leq N_c^{(k)})
\end{cases}
\end{equation}
where
\begin{equation*}
n_c^{(k)} = \lfloor L_c(2^{R_k/L_c-\lfloor R_k/L_c\rfloor} - 1)\rfloor.
\end{equation*}
Then, practically achievable data rate $\bar{R_k}$, considering finite granularity, on each subset $S_k$ is given as
\begin{align}
\bar{R_k}=&\lfloor L_c(2^{R_k/L_c-\lfloor R_k/L_c\rfloor} - 1)\rfloor \times (\lfloor R_k/L_c\rfloor + 1) \notag\\
&+ (L_c - \lfloor L_c(2^{R_k/L_c-\lfloor R_k/L_c\rfloor} - 1)\rfloor) \times \lfloor R_k/L_c\rfloor
\end{align}
Now we can optimally assign a particular modulation order to each code on different subsets. The energy,  $e_i^{(k)}$, assigned to these codes is given as

\begin{equation}\label{eik}
e_i^{(k)} = (2^{b_i^{(k)}} - 1)\frac{\Gamma}{L_c^2}N_0 \sum_{n\subset S_k} \frac{1}{|h_n|^{2}}
\end{equation}
which satisfies $\displaystyle\sum_{i} e_i^{(k)}<E_s$. This energy allocation considers null noise margin as each code receives the exact amount of energy to transmit the number of bits determined by the algorithm for a given $\Gamma$. As discussed above, $\Gamma$ is defined for a given target SER with uncoded QAM and have a constant value for all the modulation orders. In this paper we are going to deal with fixed target bit error rate (BER) instead of SER. Then taking into account the coding gains and fixed BERs, $\Gamma$ is no more constant for all the modulation orders. The above algorithm is modified to accommodate variable $\Gamma$ for different modulation orders. The exact values of the SNR gaps for all the modulation orders are stored in a predefined table and are denoted by $\Gamma_{i}^{(k)}$. These values are calculated on the basis of the selected channel coding scheme and the required system margin, which we are going to discuss in Section~\ref{theory}. For a given subset $S_k$, initially, we can take any value for $\Gamma$, say $\Gamma_{i}^{initial(k)}=1$. The closer the initial value to the exact value, the more efficient is the algorithm. $R_k$ is calculated from (\ref{eqn-1}) using $\Gamma_{i}^{initial(k)}$ while $b_i^{(k)}$ and $n_c^{(k)}$ from (\ref{bik}). Now the exact value of SNR gap, $\Gamma_{i}^{(k)}$, is taken from the table depending upon the bit vector $b_i^{(k)}$, and $e_i^{(k)}$ is calculated from (\ref{eik}) using $\Gamma_{i}^{(k)}$. Gradually bits are added in the bit vector, $b_i^{(k)}$, till $\displaystyle\sum_i e_i^{(k)}$ exceeds the PSD limit, $E_s$, and subsequently bits are removed to respect the PSD limit. We can summarize the modified approach as follows:
\begin{enumerate}
\item Calculate $R_k$, $n_c^{(k)}$ and $b_i^{(k)}$ for $\Gamma_{i}^{initial(k)}$
\item Take $\Gamma_{i}^{(k)}$, depending upon $b_i^{(k)}$
\item Calculate $e_i^{(k)}$ for $\Gamma_{i}^{(k)}$
\item Start a counter, say $count=1$
\item While$\left(\displaystyle\sum_{i} e_i^{(k)}<E_s\right)$
\begin{itemize}
\item[a)] $b_{n_c+count}^{(k)}\ +=\ 1$
\item[b)] $count\ +=\ 1$
\item[c)] update $e_i^{(k)}$
\end{itemize}
\item While$\left(\displaystyle\sum_i e_i^{(k)}>E_s\right)$
\begin{itemize}
\item[a)] $count\ -=\ 1$
\item[b)] $b_{n_c+count}^{(k)}\ -=\ 1$
\item[c)] update $e_i^{(k)}$
\end{itemize}
\end{enumerate}
\section{Theoretical Coding effects on system performance}\label{theory}
In this section, we consider the theoretical coding gain promised by the proposed concatenated channel coding scheme. In this analysis we need to deal with 2D error rates, BERs, and RS SERs, depending upon what part of the system is being considered. We use the assumptions given in~\cite{transcioffi}-\cite{confcioffi} for the sake of simplicity. Contrary to~\cite{transcioffi}-\cite{confcioffi}, all calculations are made dealing only with BERs. In these assumptions, these quantities are related by constant factors, and the 2D error rate is used as a common basis. In particular, 2D SERs are converted to BERs by multiplying by one-half. Similarly, 2D SERs are converted to RS SERs by multiplying by a constant $c$, where $c$ represents the average number of precoding sequences contributing bits to each RS symbol~\cite{transcioffi}. $P_{bit}$ denotes the required BER at the output of the overall system. From~\cite{primercioffi}, the probability of 2D symbol error in quadrature amplitude modulation is closely approximated by 
\begin{equation}\label{eqn-2}
P_{2D} \leq 4 Q\bigg[\frac{d_{min}}{2\sigma}\bigg]
\end{equation}
where $d_{min}$ is the minimum distance between QAM constellation points at the channel output, $\sigma$ is the noise variance, and $Q[.]$ represents the well-known Q-function. By using the first assumption, as discussed above, the SNR gap $\Gamma$ for a target BER of $10^{-7}$ is given as
\begin{equation}\label{eqn-3}
\Gamma = 9.8 + \gamma_{m} - \gamma_{c}~~~~(dB)
\end{equation}
where $\gamma_{m}$ is the desired margin in the system and $\gamma_{c}$, the coding gain for the proposed concatenated channel coding scheme, is given as
\begin{equation}\label{eqn-4}
\gamma_{c} = \gamma_{tc,dB} + \gamma_{rs,dB} - \gamma_{loss,dB}~~~~(dB)
\end{equation}
where $\gamma_{tc,dB}$ and $\gamma_{rs,dB}$ are the gains provided by the trellis code and the RS code respectively and $\gamma_{loss,dB}$ is the loss incurred for increasing the data rate.\\

$\gamma_{rs,dB}$: While assuming efficient interleaving to have random errors at the input of RS decoder and assuming that RS decoder does not attempt to correct the codeword if greater than $t$ errors are detected, we may relate the output RS SER, $P_{rs}$, to the input RS SER, $P_s$, by
\begin{equation}\label{eqn-5}
P_{rs} = \sum_{i=t+1}^{n} \binom{n-1}{i-1} P_s^i (1-P_s)^{n-i}.
\end{equation}
Given $P_{bit}$ and knowing that $P_{2D}=2P_{bit}$ and $P_{rs}=cP_{2D}$, we can say that $P_{rs}=2cP_{bit}$ and by iteratively solving (\ref{eqn-5}) for $P_s$, the corresponding BER at the input of RS decoder is given by
\begin{equation}\label{eqn-6}
P_{b} = \frac{P_s}{2c}
\end{equation}
and $P_b$ is the BER at the output of the demodulator, therefore an SNR gap to obtain $P_b$, $\Gamma_{rs}$, can be written as
\begin{equation}\label{eqn-7}
\Gamma_{rs} = \frac{1}{3} \Bigg(Q^{-1}\bigg[\frac{P_{b}}{2}\bigg]\Bigg)^2
\end{equation}
$\Gamma_{0,P_{bit}}$ is defined as an SNR gap required by an uncoded system to achieve $P_{bit}$, and is given as
\begin{equation}\label{eqn-8}
\Gamma_{0,P_{bit}} = \frac{1}{3} \Bigg(Q^{-1}\bigg[\frac{P_{bit}}{2}\bigg]\Bigg)^2
\end{equation}
From (\ref{eqn-8}) and (\ref{eqn-7}), $\gamma_{rs}$ can be given as
\begin{equation}\label{eqn-9}
\gamma_{rs} = \Gamma_{0,P_{bit}} - \Gamma_{rs}~~~~(dB)
\end{equation}

$\gamma_{tc,dB}$: As $P_b$ is the required BER at the input to the RS decoder and $\Gamma_{0,P_{b}}$ and $\Gamma_{tc,P_{b}}$ are the SNR gaps required by an uncoded and a Wei's 4D 16-states trellis coded system respectively to achieve $P_{b}$. Then the coding gain of a Wei's 4D 16-states trellis code can be given by
\begin{equation}\label{eqn-10}
\gamma_{tc} = \Gamma_{0,P_{b}} - \Gamma_{tc,P_{b}}~~~~(dB)
\end{equation}

$\gamma_{loss,dB}$: If $P_{tot}^*(b)$ is the minimum amount of power required to achieve the data rate $b$ as defined in~\cite{transcioffi}, the loss for the increased data rate associated with the RS code, $\gamma_{loss,dB}$, can be given as

\begin{equation}\label{eqn-12}
\gamma_{loss,dB} = P_{tot,dB}^*(\frac{nb}{k}) - P_{tot,dB}^*(b)
\end{equation}

\section{Simulation results}\label{simres}
\begin{figure}[!t]
\begin{center}
\includegraphics[scale=0.47]{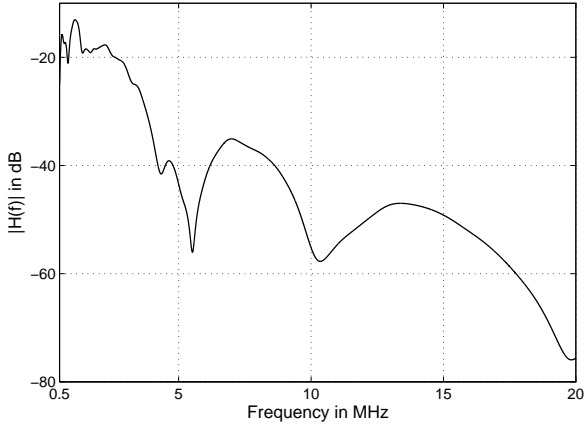} 
\caption{15-paths reference channel model for PLC~\cite{zimmermann}}
\label{fig3}
\end{center}
\end{figure}

In this section, simulation results are presented for the proposed concatenated channel coding scheme combined with the adaptive LP-DMT system. The performance of the coded adaptive LP-DMT system is compared with that of the coded DMT. The generated LP-DMT signal is composed of $N$=1024 subcarriers transmitted in the band [500;20,000]~kHz and the precoding factor, $L_c$=16. The subcarrier spacing is $19.043$~kHz. It is assumed that the synchronization and channel estimation tasks have been successfully performed. We use the multipath model for the power line channel as proposed in~\cite{zimmermann} and shown in Fig.~\ref{fig3}. The considered reference model is 110~m link 15-paths model whose frequency response is given by
\begin{equation}\label{eqn-11}
H(f) = \sum_{i=1}^{N} g_i \cdot e^{-(a_0+a_1f^k)^{d_i}} \cdot e^{-j2\pi{f\tau_i}}
\end{equation}
a result which has been widely proved in practice. The parameters of the 15-path model are listed in Table~\ref{table1}, and $\tau_i$ is the delay of path $i$. A background noise level of -110 dBm/Hz is assumed and the signal is transmitted with respect to a flat PSD of -40 dBm/Hz. Super-constellation size for Wei's 4D 16-states trellis coding is 1024 points. Results are given for a fixed target BER of $10^{-7}$.
\begin{table}[!t]
\caption{PARAMETERS OF THE 15-PATH MODEL}
\label{table1}
\begin{center}
\begin{tabular}{c|c|c|c|c|c}
\hline\hline
\multicolumn{6}{c}{\textbf{attenuation parameters}} \\
\hline
\multicolumn{2}{c|}{$k=1$} & \multicolumn{2}{c|}{$a_0=0$} & \multicolumn{2}{c}{$a_1=2.5\cdot10^{-9}$} \\
\hline\hline
\multicolumn{6}{c}{\textbf{path-parameters}} \\
\hline
$i$ & $g_i$ & $d_i (m)$ & $i$ & $g_i$ & $d_i (m)$ \\
\hline
1 & 0.029 & 90 & 9 & 0.071 & 411 \\
\hline
2 & 0.043 & 102 & 10 & -0.035 & 490 \\
\hline
3 & 0.103 & 113 & 11 & 0.065 & 567 \\
\hline
4 & -0.058 & 143 & 12 & -0.055 & 740 \\
\hline
5 & -0.045 & 148 & 13 & 0.042 & 960 \\
\hline
6 & -0.040 & 200 & 14 & -0.059 & 1130 \\
\hline
7 & 0.038 & 260 & 15 & 0.049 & 1250 \\
\hline
8 & -0.038 & 322 & & & \\
\hline\hline
\end{tabular}
\end{center}
\end{table} 

\begin{figure}[!t]
\begin{center}
\includegraphics[scale=0.47]{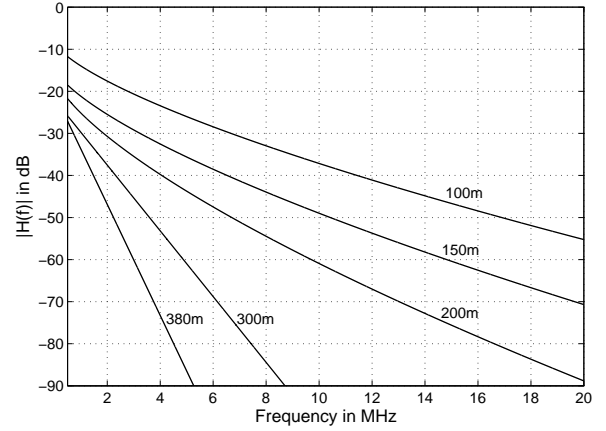} 
\caption{Length profiles of the attenuation of power line links}
\label{fig4}
\end{center}
\end{figure}

Length profile of the attenuation of power line links, i.e. neglecting the impacts of notches, as proposed in~\cite{zimmermann}, are shown in Fig.~\ref{fig4} and the corresponding parameters are listed in Table~\ref{table2}. These profiles are used to compare the performance of LP-DMT with DMT at various distances for both coded and uncoded implementations. The DMT system can be obtained by taking $L_c=1$ in the LP-DMT system. The simulations are run for a single user multiple block scenario. The proposed adaptive coded LP-DMT system can easily be extended to a multi user multiple block scenario. 
\begin{table}[!t]
\caption{ATTENUATION PARAMETERS CORRESPONDING TO THE LENGTH PROFILES}
\label{table2}
\begin{center}
\begin{tabular}{c|c|c|c|c}
\hline\hline
$class$ & $g_1$ & $a_0[m^{-1}]$ & $a_1[s/m]$ & $k$ \\
\hline\hline
$100 m$ & $1$ & $9.40\cdot10^{-3}$ & $4.20\cdot10^{-7}$ & $0.7$ \\
\hline
$150 m$ & $1$ & $1.09\cdot10^{-2}$ & $3.36\cdot10^{-7}$ & $0.7$ \\
\hline
$200 m$ & $1$ & $9.33\cdot10^{-3}$ & $3.24\cdot10^{-7}$ & $0.7$ \\
\hline
$300 m$ & $1$ & $8.40\cdot10^{-3}$ & $3.00\cdot10^{-9}$ & $1$ \\
\hline
$380 m$ & $1$ & $6.20\cdot10^{-3}$ & $4.00\cdot10^{-9}$ & $1$ \\
\hline\hline
\end{tabular}
\end{center}
\end{table}
\begin{figure}[!t]
\begin{center}
\includegraphics[scale=0.47]{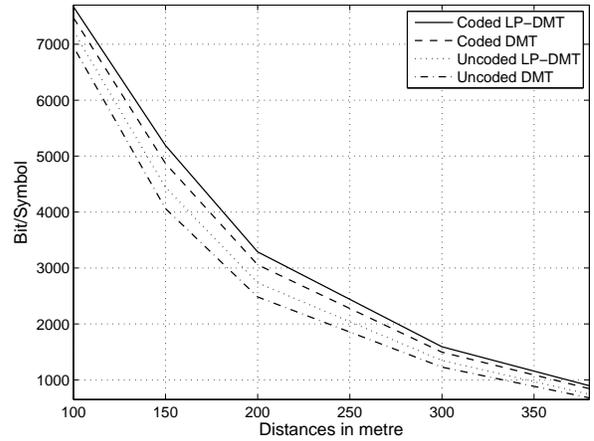} 
\caption{Achieved throughputs at various link lengths}
\label{fig5}
\end{center}
\end{figure}

\begin{figure}[!t]
\begin{center}
\includegraphics[scale=0.47]{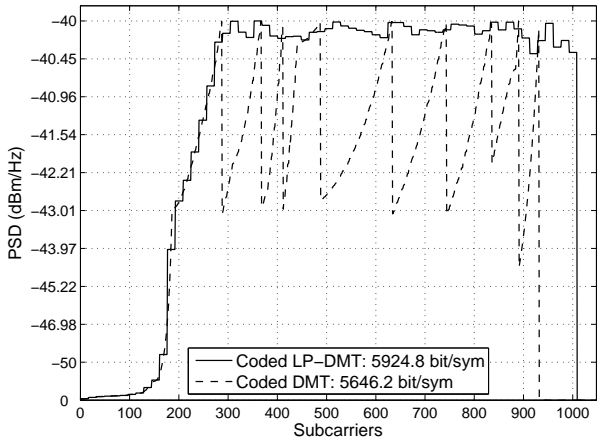} 
\caption{Energy distribution comparison of coded LP-DMT with coded DMT}
\label{fig6}
\end{center}
\end{figure}

The numbers of useful bits in each LP-DMT and DMT symbol are shown in Fig.~\ref{fig5} for both coded and uncoded implementations at various link distances, while using the bit and energy allocation algorithm as discussed in Section \ref{sysmod}. The reason for the better performance of the coded LP-DMT system is explained in Fig.~\ref{fig6}, where energy distribution of the coded LP-DMT is compared with that of the coded DMT. It is clear that the coded DMT is not fully exploiting the available energy on each subcarrier due to finite granularity and PSD constraints, while the precoding component of the coded LP-DMT system accumulates the energies of a given subset of subcarriers to transmit additional bits. Both systems respect the PSD constraint of -40 dBm/Hz as defined earlier. The coded adaptive LP-DMT system utilize more efficiently this PSD limit in comparison with the coded DMT system. Fig.~\ref{fig6} gives the minimal required energy allowing the transmission of the maximum data rate. According to the PSD mask, the residual available energy would not lead to any increase in the data rate. The spike-shaped curve of the coded DMT shows the transitions of the modulation orders (i.e. decreasing the constellation sizes) when no more energy is available to sustain the fixed target BER.
\begin{table}[!t]
\caption{THROUGHPUT COMPARISON USING 15-PATH REFERENCE MODEL OF PLC CHANNEL}
\label{table3}
\begin{center}
\begin{tabular}{c|c}
\hline\hline
system & useful bit/symbol \\
\hline\hline
coded LP-DMT & 5924.8 \\
\hline
coded DMT & 5646.2 \\
\hline
uncoded LP-DMT & 5016 \\
\hline
uncoded DMT & 4636 \\
\hline\hline
\end{tabular}
\end{center}
\end{table}

Finally, the simulations for coded and uncoded DMT and LP-DMT systems are run using the reference model of 110~m link with 15 multipaths. The results of these simulations are summarized in Table~\ref{table3}. As it is shown in Fig.~\ref{fig5} and Table~\ref{table3} that coded LP-DMT system has the highest throughput and more efficient energy utilization when compared to coded DMT (see Fig.~\ref{fig6}). Table~\ref{table3} shows that there is an improvement in the throughput of approximately 18\% and 5\% when we compare our coded LP-DMT system with uncoded LP-DMT and coded DMT respectively using 15-path reference model for power line communication for a fixed target BER of $10^{-7}$. It is worthy to mention here that 5\% is the minimum improvement achieved, due to the suboptimal subcarrier sharing method used herein, which can further be improved by a subcarrier swapping approach.
\section{Conclusion}\label{conc}
In this paper, we have investigated the resource allocation problem of an adaptive LP-DMT system taking into account the channel coding scheme. The proposed loading algorithm accommodates coding gains of the channel coding scheme in bit and energy calculations and can handle different values of the SNR gaps for different modulation orders. This bit and energy allocation algorithm can be used in combination with any channel coding scheme provided the SNR gaps of that scheme are known for all the modulation orders. A concatenated channel coding scheme, consisting of an inner Wei's 4D 16-states trellis code and an outer RS code, is proposed for the adaptive LP-DMT system in order to achieve high performance transceivers for power line communication applications. It is shown that by using a powerful but low-complexity coding scheme with the proposed algorithm, throughput of LP-DMT system can be increased significantly, further the combination of coding and precoding elements in a multicarrier power line scenario achieves higher throughput.





\begin{thebibliography}{99}
\bibitem{mallier} S.~Mallier, F.~Nouvel, J.-Y.~Baudais, D.~Gardan, and A.~Zeddam, ``Multicarrier CDMA over copper lines - comparison of performances with the ADSL system,'' in \emph{Proc IEEE Int. Workshop Electron. Design, Test, Appl.,} Jan. 2002, pp. 450-452.
\bibitem{boutros} J.~Boutros, and E.~Viterbo, ``Signal space diversity: a power- and bandwidth-efficient diversity technique for the Rayleigh fading channel,'' \emph{IEEE Trans. Inform. Theory,} vol. 44, pp. 1453-1467, Jul 1998.
\bibitem{hero} A.~O.~Hero, and T.~L.~Marzetta, ``Cutoff rate and signal design for the quasi-static Rayleigh fading space–time channel,'' \emph{IEEE Trans. Inform. Theory,} vol. 47, pp. 2400-2416, Sep. 2001.
\bibitem{rainish} D.~Rainish, ``Diversity transform for fading channels,'' \emph{IEEE Trans. Commun.,} vol. 44, pp. 1653-1661, Dec. 1996.
\bibitem{matthieu1} M.~Crussi\`ere, J.-Y.~Baudais, and J.-F.~H\'elard, ``Robust and high-bit rate communications over PLC channels: a bit-loading multi-carrier spread-spectrum solution,'' in \emph{Proc. IEEE Int. Symp. Power. Line. Commun.,} Apr 2005, pp. 37-41.
\bibitem{matthieu2} M.~Crussi\`ere, J.-Y.~Baudais, and J.-F.~H\'elard, ``Adaptive spread-spectrum multicarrier multiple-access over wirelines,'' \emph{IEEE Journal on Selected Areas in Communications,} vol. 24, no. 7, pp. 1377-1388, 2006.
3\bibitem{wei} L.F.~Wei, ``Trellis-coded modulation with multidimensional constellations'' \emph{IEEE Trans. Inform. Theory,} vol. 33, no. 4, pp. 483-501, Jul. 1987.
\bibitem{ituvdsl2} International Telecommunication Union-Telecommunication Recommendation, \emph{Very High Speed Digital Subscriber Line Transceivers 2 (VDSL2)}, ITU-T Rec.~G.993.2 Feb.~2006.
\bibitem{transcioffi} T.N.~Zogakis, J.T.~Aslanis, and J.M.~Cioffi, ``A Coded and shaped discrete multitone system,'' \emph{IEEE Trans. Commun.,} vol. 43, no. 12, pp. 2941-2949, Dec. 1995.
\bibitem{confcioffi} T.N.~Zogakis, J.T.~Aslanis, and J.M.~Cioffi, ``Analysis of a concatenated coding scheme for a discrete multitone modulation system,'' in \emph{1994 IEEE Military Communications Conf.,} pp. 433-437.
\bibitem{zimmermann} M.~Zimmermann and K.~Dostert, ``A multipath model for the powerline channel,'' \emph{IEEE Trans. Commun.,} vol. 50, no. 4, pp. 553, Apr. 2002.
\bibitem{etsivdsl} European Telecommunications Standards Institute, \emph{Very High Speed Digital Subscriber Line (VDSL)}, ETSI TS 101 270-2, Jul. 2003.
\bibitem{primercioffi} J.~Cioffi, A multicarrier primer ANSI T1E1.4/91-157, 1991, Committee contribution, Tech. Rep.
\bibitem{rheecioffi} W.~Rhee, and J.M.~Cioffi, ``Increase in capacity of multiuser OFDM system using dynamic subchannel allocation,'' in \emph{Proc. IEEE Vehicular Technology Conference (VTC-Spring'00),} vol. 2, Tokyo, Japan, May. 2000, pp. 1085-1089.
\bibitem{giannakis} Z.~Liu, Y.~Xin, and G.B.~Giannakis, ``Linear constellation precoding for OFDM with maximum multipath diversity and coding gains'' \emph{IEEE Trans. Commun.,} vol. 51, no. 3, pp. 416-426, Mar. 2003.
\bibitem{bilgiri} E.~Biglieri, ``Coding and modulation for a horrible channel,'' \emph{IEEE Commun. Mag.,} vol. 41, no. 5, pp. 92-98, May. 2003.
\end{thebibliography}
%




\end{document}